%Paper: astro-ph/9407065
%From: ksahu@ll.iac.es (Kailash C. Sahu)
%Date: Fri, 22 Jul 94 00:05:11 +0100
%Date (revised): Fri, 22 Jul 94 00:51:29 +0100

%tex format, 6 pages
\baselineskip 0.5cm
\magnification=1200
\parskip 0.3cm
\parindent=0pt
{\voffset 3.2cm
\parskip 0.2cm
\obeylines
{\centerline{\bf Stars within the Large Magellanic Cloud as potential lenses}}
{\centerline{\bf for observed microlensing events}}
\vskip 1.2cm
{\centerline{ Kailash C. Sahu}}
{\centerline{ Instituto de Astrofisica de Canarias}}
{\centerline{ 38200 La Laguna, Tenerife, Spain}}
{\centerline{\bf e-mail: ksahu@iac.es}}
\vskip 1.5cm
{\centerline{Accepted for publication as a}}
{\centerline{ {\bf Letter to Nature}}}
\vfill \eject}

{\bf Abstract: \par
Massive Compact objects in the halo, known as MACHOs, have been postulated as
the origin of a substantial fraction of `dark matter' known to exist in the
haloes of galaxies$^{1,2}$. Paczy\'nski$^3$ has suggested that it might
possible to detect these low-luminosity objects by their potential to
act as gravitational lenses, causing a characteristic brightening when
they cross the path of light from a star in a nearby galaxy.
very recently, two groups reported possible detections of microlensing of
stars in the Large Magellanic Cloud (LMC)$^{4,5}$. Here I show that
microlensing by stars within the LMC itself can account for the observed
events. It is further shown that if stars within the LMC are the lenses,
the observed light curve can differ from the light curve due to a
galactic lens even at relatively low magnifications.
This provides a possibility of distinguishing between the galactic lenses and
the LMC lenses. For a given number of monitored stars,
the LMC induced events should be strongly concentrated towards the
central region of the LMC, while the galactic events should be
uniformly spread over the whole area of the LMC, so the two can also
be distinguished statistically.
}

Based on a suggestion by Paczy\'nski$^3$, two groups involved
in the monitoring of a  few million stars in the LMC for more than a year,
have reported the detection of three microlensing events as evidence for
the presence of dark objects in the Galactic halo$^{4,5}$.
However, this has also been the cause for some concern:  if these events are
caused by such objects in the halo, it would lead to some problems in
terms of stellar evolution and the theory of galaxy formation$^6$.
Recent work on deuterium abundance derived from the observations
of a quasar suggests that the dark halos of galaxies may well be
nonbaryonic$^7$.
Furthermore, the rate of
microlensing events as observed seems to be lower than expected from the
dark halo, and it is important to estimate the rate expected from the stars
that are known to exist in our galaxy and in the LMC.  The analyses for the
galactic stars and the LMC halo have been done$^{8,9}$, but the importance
of the LMC stars
has been overlooked so far.  This paper presents the estimate of the rate of
microlensing events to be expected from the LMC stars acting as gravitational
lenses.

To calculate the probability of the microlensing being caused by a star in
the LMC itself, we need to know the stellar mass density in the LMC.
Let us first confine ourselves to the bar of the LMC. From the surface
luminosity, it is estimated that the observed luminosity
of the bar is about 10 to 12 \% of the total observed luminosity of the LMC
in the optical wavelengths$^{10,11}$.
To calculate the extinction,
I have used the IRAS 100 $\mu$m maps$^{12,13}$, the background galaxy
counts$^{14}$, and the observations of
the most reddened stars in the region$^{15}$, which give a consistent
value of 1.5 magnitudes in V.
To see how the extinction affects the mass, let us assume that
the extinction is uniform in depth, and let $d$ be the total depth.
Thus for any line of sight, if A$_v$ is the total extinction in the line of
sight
$${L_{obs}\over{L_{true}}} = {1\over d} \int_0^d{2.5^{-A_vl/d} dl}
 =  {1 - e^{-0.916 A_v}\over{0.916 A_v}}\eqno (1)$$
 Using  A$_v$=1.5 magnitudes for the bar, and A$_v$=0.3 to 0.4 magnitudes for
the region outside, it is easy to show that the true luminosity of
the bar is about 14 to 18\% of the total luminosity of the LMC. The total
mass of the LMC, as determined from
various methods, ranges from 6 x 10$^9$ to 15 x 10$^9$ M$_\odot^{10,16}$.
Assuming that the mass to light ratio in
the bar and the outer parts are the same, we can take the  mass of the bar as
2 x 10$^9$ M$_\odot$. Considering the fact that the LMC
is gas poor  and only about 5\% of the LMC
mass is neutral hydrogen$^{16-18}$,  we will neglect the contribution
of gas to the mass and assume that the entire mass is made up of stars.

To calculate the probability of microlensing, let us assume that there are
N$_{tot}$ stars being monitored.
To see the effect of extinction on the number
of monitored stars at various depths, we note that
the limiting magnitude of the current surveys are about 20 to 21, which at the
distance of the LMC, corresponds to an absolute magnitude of 1.5 to 2.5. So
the
magnitude of the stars that can be observed at the near side is 1.5 to 2.5
while the limiting magnitude  at the far end is 1.5 magnitudes brighter.
If the distribution of stars among different spectral types is assumed to
be similar to what is observed in the solar neighborhood, then the difference
in the monitored number of stars per unit depth from
front end to the back is about three$^{19}$.
(As it turns out,
the effect of extinction is small.
If the extinction is 3 magnitudes instead of 1.5,
the net probability decreases only by less than 50\%.)

Hence, assuming the extinction to be uniform in depth, we can express the
observed number of stars at any layer $dl$, at a depth of $l$, as
$$N_{obs}(l) = {N_l} 3^{-{l\over d}} dl \eqno (2)$$
where $N_l$ is the observed number of stars per unit depth in absence of
extinction.
%i.e, the observed number of stars per unit depth at the near side.
If $N_{tot}$ is the total number of stars being monitored, then
$$N_{tot} = \int_0^d{N_l 3^{-{l\over d}} dl} = 0.6 N_l d \eqno (3)$$
Substituting eq. 3 in eq. 2, we get
$$N_{obs}(l) \simeq  {N_{tot}\over 0.6d}  {3^{-{l\over d}} dl}\eqno (4)$$
The fraction of area covered by the Einstein rings af all  the individual
stars lying in the front of this layer can be expressed as
$$A_f(l) = \int_0^l{\pi R_E^2(l) n(l)  dl} \eqno (5) $$
where $n(l)$ is the stellar number density at depth $l$ and R$_E$ is the
Einstein radius. Note that  $n = \rho /m$
and $R_E^2 \propto m$ where $\rho$ is the stellar mass density and $m$ is
the mass of the star. Thus,
assuming the stellar
density to be uniform with depth
 $$A_f(l) =  {2 \pi G \rho l^2 \over{c^2}}\eqno (6) $$
which is thus independent of the mass distribution.
 From Eq. 4 and 6, the optical depth to microlensing by stars within the bar
can be expressed as
$$ \tau_{bar} = {1\over N_{tot}}\int_0^d{N_{obs}(l) A_f(l) dl} \simeq
{0.5 \pi G \rho_{bar}  d^2 \over c^2} \eqno (7)$$
For the value of $\rho$, we can substitute,
$$\rho_{bar} = {M  \over {L W d}} \eqno (8)$$
where L, W, d and M are the length, width, depth and mass of the bar
respectively.
Assuming L=3000pc and W=d=600pc$^{10,20,21}$,
and substituting eq.8  in eq.7,  we get,
$$\tau_{bar} \simeq 5 \times 10^{-8}, \eqno (9)$$
with some uncertainties due to the uncertainties in the size and mass
of the bar.
Carrying out an identical analysis for the region outside the bar
it is easy to see that, if the depth of the LMC disk is
between 100 to 300pc$^{22}$, the optical depth in the region outside the bar
is about 4 to 12 times smaller.

The MACHO event  lies in the central region of the
bar.
Keeping in mind the uncertainties involved in
the estimation of optical depth on the basis of a single event$^{8,23}$, we
can only say that it
seems to be well below the optical depth of
$5 \times 10^{-7}$  expected from a dark halo made up entirely of
MACHOs, and is consistent with  the optical
depth calculated above.
In the case of EROS events, one lies in the outer region of the bar,  the
other is far from the bar,
and the optical depth has been estimated to be  higher
($\sim 2 \times 10^{-7}$, Ref. 8).
But Gould et al.$^8$ also suggest that there may be some systematic effects
and the detection efficiencies may be uncertain. The situation will be
clearer as more events are observed, particularly events with higher
magnifications (see below), and we must wait for more events to be
observed before making any direct comparison with the calculated
optical depth.

If the microlensing is indeed caused by the LMC lenses,
then one important consequence is that the sources can be resolved at much
smaller magnifications. Let us see how.

The amplification sharply rises  when the
lensing object comes close to being perfectly aligned with the source.
The physical reason for this is that when the lensing object and
the source are perfectly aligned, the image becomes a ring (also
called the Einstein ring) instead of
two separate images.
 [For details, see ref. 3]. In the case of lensing by MACHOs,
$R_E$ is of the order of 1.2 $\times$ 10$^{14}$ $\sqrt{M\over M_\odot}$ cm,
 which, at the source plane, is
$\sim$10$^{15}\sqrt{M\over M_\odot}$ cm.
If the lensing is caused by the objects in the disk of our galaxy, $R_E$
projected onto the source plane is $>10^{15}\sqrt{M\over M_\odot}$.
But if the lens is in the LMC, for a
typical distance D of about 100pc between the source and the lens,
R$_E \simeq  10^{13} \sqrt{M\over
M_\odot}$ cm, which is at least 2 orders of magnitude smaller.
As an example, let us consider the source to be a red giant, and a  0.5
M$_\odot$ lens. In order that at least a part of the source
is perfectly aligned with the lensing object in this case, the impact
parameter $u$  has to be $^< \hskip -0.2cm _\sim$0.001 (i.e. magnification
$^> \hskip -0.2cm _\sim$ 1000) for a halo or disk
lens whereas it has to be  $^< \hskip -0.2cm _\sim$0.1 (i.e. magnification
$^> \hskip -0.2cm _\sim$ 10) for an LMC lens, which will make the light curve
different from that
of a point source$^{24-26}$. Considering
the fact that the events are expected
to be uniformly distributed in $u$, this provides a possibility to
distinguish between the halo lenses and the LMC lenses as more events are
observed.

As is clear from the optical depth estimates,
for a given number of monitored stars, the LMC induced events should be
strongly concentrated
towards the central region of the LMC, while the galactic events (whether
by disk or halo lenses) should be
uniformly spread over the whole area of the LMC. So
the two can be distinguished statistically in most cases and,
as described earlier, individually in some cases from the light curves.

{\it Acknowledgements}   Harvey Butcher, Paul Hodge and Bengt
Westerlund promptly replied to my queries which were  helpful in clarifying
some  doubts regarding the structure
of the LMC. Some details of the observational program
provided by Ken Freeman are gratefully acknowledged.
It is a pleasure to thank  the referees
Bohdan Paczy\'nski, Alain Milsztajn and two anonymous referees for
their comments, which were helpful in improving the paper.
\vfill \eject

{\parskip 0.05cm
{\bf References}\par
1. Begeman, K., Ph.D. thesis, Groningen University (1987)\par
2. Fich, M., Blitz, L. \& Stark, A., Astrophys. J., {\bf 342}, 272-284
(1989)\par
3. Paczy\'nski, B.,  Astrophys. J., {\bf 304}, 1-5 (1986)\par
4. Alcock, C. et al., Nature, {\bf 365}, 621-623 (1993)\par
5. Aubourg et al.,  Nature, {\bf 365}, 623-625 (1993)\par
6. Hogan C.J., Nature, 365, 602-603 (1993)\par
7. Songaila, L.L., Cowie, C.J., Hogan, C.J. \& Rugers, M., Nature, {\bf 368},
599-604 (1994)\par
8. Gould, A., Miralda-Escud\'e, J. \& Bahcall, J.N., Astrophys. J. Lett.,
{\bf 423}, L105-L108 (1994)\par
9. Gould, A., Astrophys. J., {\bf 404}, 451-454 (1993)\par
10. De Vaucouleurs, G. \& Freeman, K.C., Vistas Astro. {\bf 14}, 163-294
(1973)\par
11. Bothum, G.D. \& Thompson, I.B., Astro. J., {\bf 96}, 877-883 (1988)\par
12. Schwering P.B.W. \& Israel, F.P., Atlas and Catalog of Infrared
Sources in Magellanic Clouds (Kluwer Academic Publishers 1990) \par
13. Laureijs R., Ph.D. thesis, Groningen University, 85 (1989)\par
14. Hodge, P., IAU Symp. 148  ``The Magellanic Clouds", (eds
Haynes, R. \& Milne, D.) 57-62 (Kluwer Academic Publishers 1991). \par
15. Isserstedt, J.,  Astro. Astrophys. {\bf 41}, 175-182 (1975)\par
16. Westerlund, B., Astro. Astrphys. Rev., {\bf 2}, 29-78 (1990)\par
17. Israel, F.P. \& de Graauw, Th., IAU Symp. 148 ``The Magellanic Clouds",
(eds
Haynes, R. \& Milne, D.) 45-52 (Kluwer Academic Publishers 1991). \par
18. Rohlfs, K., Kreitschmann, J., Siegman, B.C. \& Feitzinger, J.V.,
Astro. Astrophys. {\bf 137}, 343-357 (1984)\par
19. Allen, C.W.,  {\it Astrophysical Quantities}, 3rd Edition, (the
Athlone Press, 1973)\par
20. Binney J. \& Tremaine, S., Galactic Dynamics (Princeton University
Press, 1987)\par
%Einstein, A., 1936, Science, {\bf 84}, 506 \par
21. Westerlund, B.,  IAU Symp. 148 ``The Magellanic Clouds", (eds
Haynes, R. \& Milne, D.) 15-23 (Kluwer Academic Publishers 1991). \par
22. Feast, M., ``Recent Developments in Magellanic Cloud Research", (eds
de Boer, K.S., Spite, F. \& Stasinska, G.) 75-86 (Observatoire de Paris,
1989). \par
23. Gould, A., Astrophys. J. Lett., {\bf 421}, L71-L74 (1994)\par
24. Schneider, P., Ehlers, J. \& Falco, E.E.,  ``Gravitational Lensing",
(Springer-Verlag, 1992).\par
25. Nemiroff, R.J. \& Wickramsinghe, W.A.D.T., Astrophys. J. Lett.,
{\bf 424}, L21-L23 (1994)\par
26. Sahu, K.C., Pub. Astron. Soc. Pacific (in the press)\par
}
\vfill \eject
\bye